\documentclass[sigplan,10pt]{acmart}
\renewcommand\footnotetextcopyrightpermission[1]{}
\AtBeginDocument{%
  }

\setcopyright{acmlicensed}
\copyrightyear{2018}
\acmYear{2018}
\acmDOI{XXXXXXX.XXXXXXX}
\acmConference[Conference acronym 'XX]{Make sure to enter the correct
  conference title from your rights confirmation email}{June 03--05,
  2018}{Woodstock, NY}
\acmISBN{978-1-4503-XXXX-X/2018/06}

\usepackage{xspace}
\usepackage{booktabs}
\usepackage{multirow}
\usepackage{xcolor}
\usepackage{amsmath}
\usepackage{listings}
\usepackage{xcolor}
\usepackage{algorithm}
\usepackage{algpseudocode}
\usepackage{amsmath}

\definecolor{codegray}{rgb}{0.5,0.5,0.5}
\definecolor{backcolour}{rgb}{0.98,0.98,0.98}
\usepackage{enumitem}
\lstdefinestyle{jsonstyle}{
    backgroundcolor=\color{backcolour},   
    commentstyle=\color{codegray},
    keywordstyle=\bfseries,
    numberstyle=\tiny\color{codegray},
    stringstyle=\ttfamily,
    basicstyle=\ttfamily\footnotesize,
    breakatwhitespace=false,         
    breaklines=true,                 
    captionpos=b,                    
    keepspaces=true,                 
    numbers=left,                    
    numbersep=5pt,                  
    showspaces=false,                
    showstringspaces=false,
    showtabs=false,                  
    tabsize=2,
    frame=single, 
    rulecolor=\color{black!30}
}
\begin{document}
\newcommand{\system}{\texttt{MorphSched}\xspace}
\title{\system: Workload-Aware Adaptive Scheduler Generation for Linux}

\author{Yifan Zhang}
\email{12451018@zju.edu.cn}
\affiliation{%
  \institution{Zhejiang University}
  \city{Hangzhou}
  \country{China}
}

\author{Hengxuan Lou}
\email{12560049@zju.edu.cn}
\affiliation{%
  \institution{Zhejiang University}
  \city{Hangzhou}
  \country{China}
}

\author{Xinkui Zhao}
\authornote{Corresponding author.}
\email{zhaoxinkui@zju.edu.cn}
\affiliation{%
  \institution{Zhejiang University}
  \city{Hangzhou}
  \country{China}
}

\author{Rengrong Xiong}
\email{12651002@zju.edu.cn}
\affiliation{%
  \institution{Zhejiang University}
  \city{Hangzhou}
  \country{China}
}

\author{Naibo Wang}
\email{wangnaibo@zju.edu.cn}
\affiliation{%
  \institution{Zhejiang University}
  \city{Hangzhou}
  \country{China}
}

\author{Chang Liu}
\email{liuchangntu@gmail.com}
\affiliation{%
  \institution{Zhejiang University}
  \city{Hangzhou}
  \country{China}
}

\author{Guanjie Cheng}
\email{chengguanjie@zju.edu.cn}
\affiliation{%
  \institution{Zhejiang University}
  \city{Hangzhou}
  \country{China}
}

\author{Shuiguang Deng}
\email{dengsg@zju.edu.cn}
\affiliation{%
  \institution{Zhejiang University}
  \city{Hangzhou}
  \country{China}
}

\author{Jianwei Yin}
\email{zjuyjw@cs.zju.edu.cn}
\affiliation{%
  \institution{Zhejiang University}
  \city{Hangzhou}
  \country{China}
}

\renewcommand{\shortauthors}{Zhang et al.}


\begin{abstract}
Modern Linux scheduling increasingly requires workload-specific policies to meet the dynamic performance needs of heterogeneous applications. The recent \texttt{sched\_ext} framework enables custom BPF schedulers, but building practical policies still requires coordinating Rust user-space control logic, BPF scheduling callbacks, generated user-kernel interfaces, and strict eBPF verifier constraints. This paper presents \system{}, an LLM-driven system for synthesizing complete Rust+BPF \texttt{sched\_ext} schedulers.

\system{} profiles target workloads under Linux EEVDF to construct dynamic, workload-specific optimization objectives, and uses a scaffold-constrained generation pipeline to translate scheduling intent into deployable scheduler implementations. To improve synthesis reliability, \system{} separates implementation repair from strategy search: verifier, build, and runtime failures are handled through a DAG-structured repair process, while performance feedback is accumulated in an evolutionary memory pool to refine high-level scheduling strategies across iterations.

We evaluate \system{} across 120 machine/workload/interferer configurations on a real Linux kernel. The generated schedulers compile, pass eBPF verification, load successfully, and remain stable throughout our benchmark runs without manual edits to generated scheduler source code. Across these configurations, \system{} improves the primary foreground metric over EEVDF by 26.5\% on average and up to 87.3\%. In selected comparisons, the generated schedulers are also competitive with representative hand-written \texttt{sched\_ext} schedulers. These results show that scaffold-constrained, system-grounded LLM generation can be a practical tool for exploring workload-specific Linux scheduling policies.
\end{abstract}



\received{20 February 2007}
\received[revised]{12 March 2009}
\received[accepted]{5 June 2009}
\settopmatter{printfolios=true}

\maketitle

\section{Introduction}

Operating system schedulers serve as a cornerstone of modern OS design, providing the fundamental abstraction for managing and allocating CPU resources across diverse computational tasks~\cite{peterson1985operating, tanenbaum1997operating, weiser1991computer, chitralekha2021survey,robert2018linux}. The architecture of a scheduler is dictated by a complex trade-off between two compelling forces: the drive for maximal hardware utilization in data centers, and the specific performance guarantees required by heterogeneous workloads~\cite{li2007efficient,goyal1996hierarchical}. This tension drives the continuous evolution of kernel scheduling policies, spurring the development of Linux's default CFS~\cite{linuxcfsdoc, bouron2018battle}, EEVDF~\cite{linuxeevdfdoc, gough2026comparative} scheduler alongside various bespoke scheduling strategies tailored for specific business logic and mixed-deployment scenarios~\cite{wang2025mixture,fried2020caladan}. Consequently, kernel and system developers find themselves trapped in an endless cycle of manually implementing, fine-tuning, and debugging workload-specific scheduling algorithms just to keep pace with innovation amid increasingly complex hardware and software ecosystems.

To accelerate this scheduling evolution, the Linux kernel has recently introduced the \texttt{sched\_ext} framework~\cite{linuxschedextdoc}, which aims to enable rapid scheduler experimentation and customization via pluggable eBPF dispatch mechanisms. However, this promising shift comes at a steep price, as implementing a production-ready scheduling policy within this new paradigm remains notoriously difficult. To quantify this challenge, we conduct a comprehensive study of existing \texttt{sched\_ext} schedulers and their development workflows(See Section~\ref{sec:moti}). Our analysis shows that the effort required to safely orchestrate, stabilize, and maintain a \texttt{sched\_ext} policy far outweighs the algorithmic design itself.

At the same time, large language models (LLMs) have become increasingly capable at generating nontrivial systems code. This raises a natural question: can we use LLMs to synthesize Linux schedulers? More specifically, given a workload can an LLM generate a complete production-ready \texttt{sched\_ext} scheduler?

This motivates us to explore a new paradigm for operating system scheduler design and development, agent-driven generative schedulers, which leverages the autonomous planning, code intelligence, and reasoning capabilities of LLM-based agents. Instead of relying on manual efforts, this paradigm empowers intelligent agents to automatically synthesize, verify, and seamlessly evolve highly customized scheduling policies, achieving rapid feature iteration and production-ready robustness tailored to specific application workloads.

However, synthesizing practical \texttt{sched\_ext} policies via LLMs extends far beyond unconstrained code generation; it is bottlenecked by three fundamental systems challenges. First, without dynamic baselines to bridge the semantic gap, agents operate blindly, prone to over-optimizing single metrics at the expense of overall system liveness. Second, viable policies demand rigid cross-language orchestration between kernel-space BPF callbacks and Rust user-space control planes, where monolithic generation inevitably triggers cascading structural mismatches. Finally, standard linear repair loops inherently induce ``defensive accumulation''---stacking redundant logic that bloats the latency-critical scheduling hot-path until it violates strict eBPF verifier bounds. To navigate this hostile synthesis environment, we introduce \system{}...

This paper presents \system{}, an autonomous system for generating Linux \texttt{sched\_ext} schedulers via large language models. To bridge the semantic gap between hardware telemetry and scheduling intents, \system profiles the target workload under Linux EEVDF to formulate a dynamic optimization objective. To conquer the cross-language structural complexity of \texttt{sched\_ext}, \system{} abandons fragile monolithic generation in favor of a contract-driven progressive architecture. It first translates strategies into a language-agnostic Intermediate Representation (IR) and deploys parallel LLM threads to realize user-space and kernel-space components strictly within this shared constraint. Each generated scheduler is compiled, verified, loaded, executed, and benchmarked. Rather than relying on naive linear feedback, \system{} navigates the high-dimensional search space through a decoupled dual-level optimization engine: it utilizes a DAG-structured semantic search to systematically bypass strict eBPF verifier bounds, and an evolutionary memory pool to recombine and refine high-level scheduling strategies without context drift.  

We evaluate \system{} on 120 configurations across diverse workloads—spanning both isolated executions and heavily-interfered colocated deployments—across five machine settings. All reported schedulers are generated automatically without manual source-code edits. Across these configurations, MorphSched improves the primary foreground metric over EEVDF by 26.5\% on average (and up to 87.3\%), demonstrating that autonomously generated, contract-constrained policies can consistently match or outperform expert-crafted, hand-written \texttt{sched\_ext} schedulers.  

This paper makes the following contributions:
\begin{itemize}
    \item We characterize the engineering complexity of
    \texttt{sched\_ext} scheduler development, showing that practical
    policies require coordinated Rust user-space logic, BPF callbacks,
    dispatch queues, statistics, and generated interfaces.

    \item We design \system{} to synthesize policies driven by dynamic EEVDF baselines. It ensures user-kernel consistency via Contract-Driven Progressive Lowering, and navigates strict eBPF bounds using a dual-level search composed of DAG-based repair and evolutionary recombination.

    \item We evaluate \system{} on real-world workloads, showing that fully automated generation can synthesize robust schedulers that consistently improve performance over both the Linux baseline and hand-written expert policies.  
\end{itemize}

\section{Background}

\subsection{Linux Scheduling Evolution}
Linux schedulers multiplex runnable tasks across CPU cores while balancing fairness, responsiveness, throughput, and system stability. Introduced in Linux 2.6.23 (2007), the Completely Fair Scheduler (CFS) served as the default timesharing scheduling algorithm in the Linux kernel for over 15 years. CFS achieves fairness among competing tasks by tracking their virtual runtime (\textit{vruntime}) and utilizing a red-black tree to always select the task with the minimum \textit{vruntime} for execution. While highly effective for generic workloads, CFS often struggles to provide tight latency guarantees for mixed-deployment scenarios, where latency-sensitive foreground services are co-located with throughput-oriented background jobs. To alleviate this, the Linux kernel recently transitioned to the Earliest Eligible Virtual Deadline First (EEVDF) scheduler in version 6.6 (late 2023). Building upon Peter Zijlstra's implementation of the EEVDF theorem, it introduces the concepts of virtual lag and virtual deadlines to allocate CPU time slices fluidly, allowing lagging tasks to preempt others earlier. The long interval between CFS and EEVDF highlights the conservative nature of mainline scheduler evolution: changes to the default Linux scheduler must be safe, general, and robust across a wide range of workloads and hardware platforms.

\subsection{\texttt{sched\_ext}: Extensible Scheduling in Linux}

Linux \texttt{sched\_ext} is an extensible scheduler framework that allows
custom scheduling policies to be implemented as eBPF \texttt{struct\_ops}
programs and loaded at runtime~\cite{linuxschedextdoc}. Instead of
modifying the core kernel scheduler, developers implement a set of BPF
callbacks that are invoked from the scheduling path. For example,
\texttt{select\_cpu} can influence CPU placement on wakeup,
\texttt{enqueue} controls how runnable tasks enter scheduler-managed
queues, and \texttt{dispatch} moves tasks from dispatch queues to idle
CPUs.

\begin{figure}[t]
    \centering
    \includegraphics[width=1\linewidth]{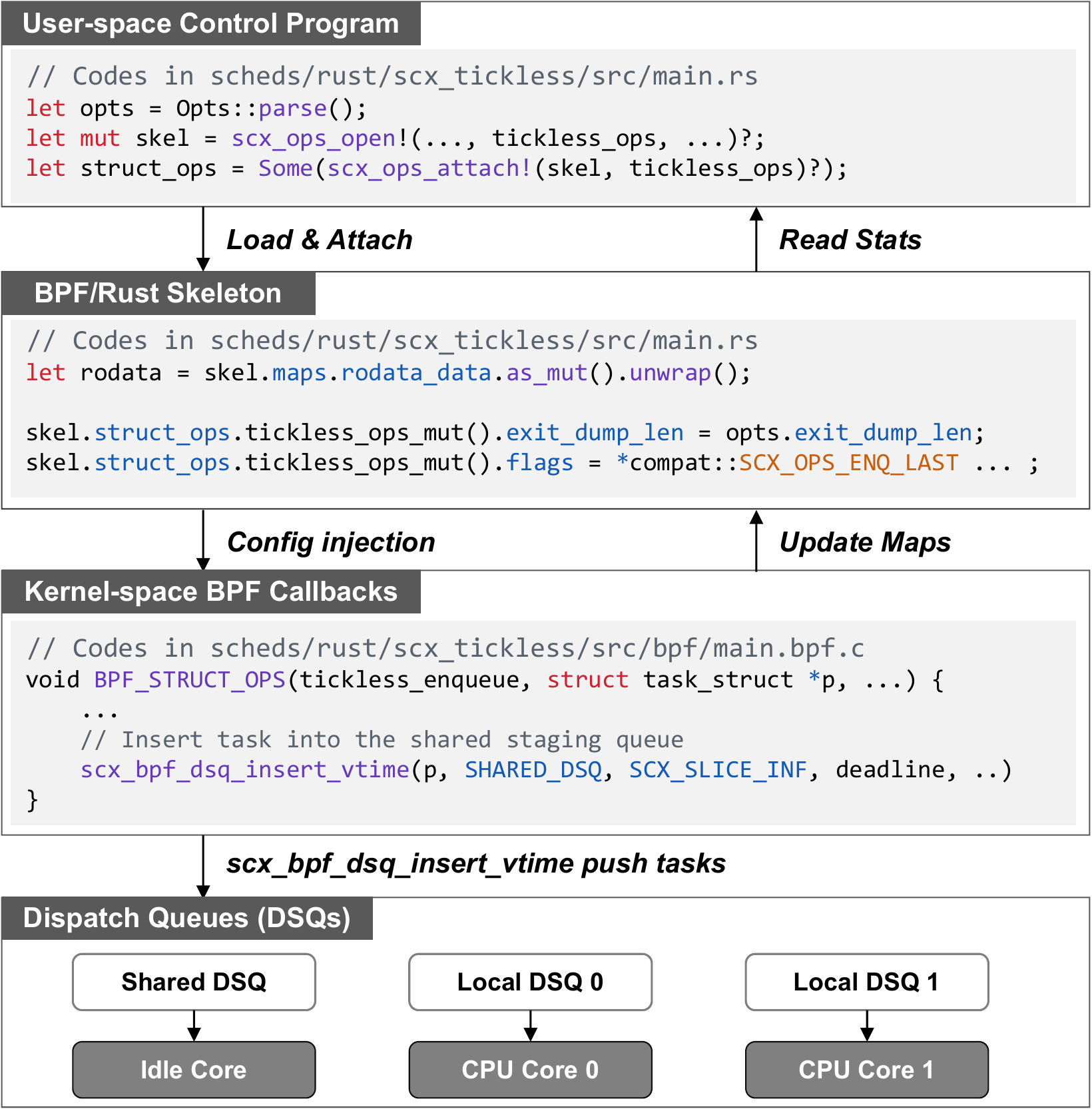}
    \caption{Representative Rust+BPF structure of a
    \texttt{sched\_ext} scheduler, illustrated using a tickless-style
    policy. User-space control logic loads and attaches the BPF
    scheduler, BPF callbacks make scheduling decisions, and dispatch
    queues (DSQs) mediate task flow to CPUs.}
    \label{fig:scx_arch}
\end{figure}

Figure~\ref{fig:scx_arch} shows the typical structure of a
\texttt{sched\_ext} scheduler. The BPF component implements scheduling
callbacks and may use built-in or user-defined dispatch queues (DSQs) to
stage runnable tasks before they are consumed by CPUs. The user-space
component, commonly written in Rust or C, loads the BPF program, sets
configuration parameters, attaches the scheduler, reads statistics, and
handles scheduler exits. Generated skeleton interfaces connect the
user-space and BPF components by exposing maps, read-only data, and
\texttt{struct\_ops} definitions across the user/kernel boundary.

\texttt{sched\_ext} also provides safety and observability mechanisms.
The kernel can reject invalid BPF programs through compilation,
verification, or load-time checks, and the framework can exit a scheduler
when it detects unsafe runtime behavior such as stalled tasks. These
signals make \texttt{sched\_ext} suitable for iterative scheduler
experimentation: a candidate policy can be generated, compiled, loaded,
observed, and rolled back without replacing the default kernel scheduler.

\section{Motivation and Key Observations}
\label{sec:moti}

We motivate \system{} with three observations about generating Linux \texttt{sched\_ext} schedulers: practical policies are cross-language systems programs, scheduler objectives are workload-coupled, and generation failures occur at different levels.

\subsection{Objective Semantic Gap}
Unlike standard standalone software, an operating system scheduler possesses no intrinsic performance metric of its own. Its efficacy is defined by how well it accommodates the dynamic runtime characteristics of the applications running above it. Consequently, automated scheduler synthesis demands injecting much more than just functional correctness constraints. This creates a profound semantic gap. Forcing an LLM to simultaneously reason about exact workload behaviors and synthesize fine-grained eBPF primitives within a monolithic prompt inherently triggers cognitive overload.

\subsection{Cross-Language Complexity of \texttt{sched\_ext} Policies}

A \texttt{sched\_ext} scheduler is not a single policy function. It spans
user-space control code, kernel-space BPF callbacks, dispatch queues,
statistics, and generated skeleton interfaces. These components must
remain consistent across language and user/kernel boundaries: Rust-side
skeleton fields must match BPF symbols, statistics must match exported
BPF variables, \texttt{struct\_ops} identifiers must match the attach
path, and helper or kfunc signatures must match the running kernel.

\begin{figure*}[t]
    \centering
    \includegraphics[width=1.0\linewidth]{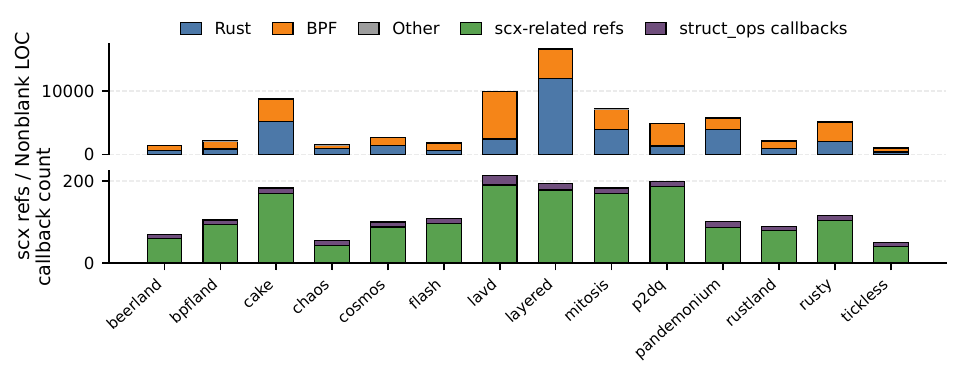}
    \caption{Engineering footprint of 14 upstream \texttt{scx}
    schedulers. Top: non-blank LOC split by Rust, BPF, and other source
    files. Bottom: number of \texttt{sched\_ext}-related references and
    implemented \texttt{struct\_ops} callbacks. Existing policies require
    coordinated Rust and BPF components rather than a single scheduling
    function.}
    \label{fig:complexity}
\end{figure*}

To quantify this footprint, we analyze 14 upstream \texttt{scx} schedulers from the \texttt{sched-ext/scx} repository~\cite{scxrepo}.
As shown in Figure~\ref{fig:complexity}, these schedulers contain 159
source files and 71,278 non-blank LOC, split almost evenly between Rust
(51.6\%) and BPF (47.9\%). Individual policies can be large:
\texttt{scx\_layered} exceeds 16K non-blank LOC. They also use many
\texttt{sched\_ext} APIs and callbacks; a single scheduler contains up to
190 \texttt{sched\_ext}-related references and up to 24 implemented
\texttt{struct\_ops} callbacks. These measurements show that automated
scheduler generation must handle complete Rust+BPF projects rather than
isolated callback snippets.

\subsection{Implementation Traps vs. Strategy Failures}

Generated schedulers exhibit two fundamentally distinct failure modes: implementation-level and strategy-level. Implementation-level failures manifest during compilation, skeleton integration, or eBPF verification. Standard generative models rely on linear conversational repair~\cite{wu2026runtime}, which inherently drives them toward ``defensive accumulation''---stacking bloated, redundant conditional logic to blindly satisfy static verifier constraints. In the latency-critical scheduling hot-path, this additive modification is fatal, degrading context-switch performance. Yet, discarding a structurally sound scheduling idea simply because the agent fell into a syntactic repair loop wastes valuable algorithmic insights. This demands a repair mechanism that can rescue the underlying intent without succumbing to agent-induced code bloat. Conversely, strategy-level failures occur when a generated candidate compiles, passes strict verification, and executes safely, but fails to improve the workload's performance objective. In these scenarios, further syntactic optimization is futile; the macroscopic scheduling direction itself must pivot.

\subsection{Implications}

These observations collectively imply that autonomous kernel-level scheduler synthesis cannot be framed as an unconstrained, snippet-level code generation task. Instead, any viable synthesis framework must adhere to three core architectural principles: first, it must ground its optimization targets in dynamic, workload-coupled environment rather than static SLOs to prevent futile optimization loops; second, it must enforce a rigid cross-language structural scaffold to insulate the generative agent from project layout chaos and ensure user/kernel skeleton consistency; and third, it must decouple the search space into a bifurcated feedback loop—applying localized code repairs for structural implementation errors while reserving macro-level strategy refreshes for safe but algorithmically ineffective policies.

\section{Design}

\begin{figure}[t]
    \centering
    \includegraphics[width=1.0\linewidth]{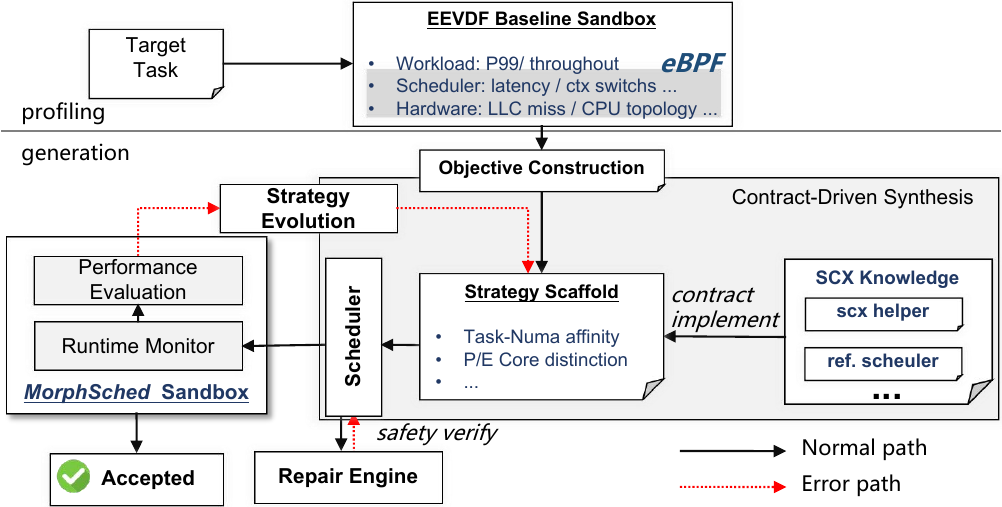}
    \caption{Architecture of \system{}.}
    \label{fig:pipeline}
\end{figure}

Figure~\ref{fig:pipeline} illustrates the overall architecture and adaptive workflow of \system{}. Driven by this pipelined framework, the remainder of this section details our concrete system mechanics. Specifically, we first present our methodology for capturing dynamic execution telemetry and performing workload-specific objective construction (Sec.~\ref{subsec:objective_construction}). We then describe the scaffold-driven synthesis engine that translates high-level scheduling strategies into full-stack scheduler code (Sec.~\ref{subsec:synthesis}). Finally, we elaborate on the hierarchical validation sandbox and the decoupled dual-level feedback loops that continuously optimize and repair the synthesized policy (Sec.~\ref{subsec:verification_and_feedback}).

\subsection{Workload Objective Construction}
\label{subsec:objective_construction}

A fundamental challenge in autonomous scheduler synthesis is defining a reliable convergence condition for the LLM agent. Prompting a model with vague qualitative goals (e.g.,  ``improve Nginx performance'') provides no quantifiable feedback, whereas enforcing hard-coded absolute Service Level Objectives (SLOs, e.g., ``tail latency $\leq$ 5ms'') is inherently brittle. Absolute SLOs ignore hardware heterogeneity and the highly variable severity of colocation interference. If a static target is physically unachievable on a given node, the LLM is inevitably driven into an infinite, futile repair loop, attempting to fix correct code to achieve an impossible physical constraint.

To bridge this semantic gap, \system{} abandons static thresholds in favor of dynamic, workload-coupled baselines. Before synthesis begins, the framework profiles the specific target-interferer pair under the default Linux EEVDF scheduler to establish a contextual ground truth. For latency-sensitive services (e.g., Nginx, Redis), let $L_{\mathrm{base}}$ denote the baseline p99 latency under EEVDF. \system{} constructs the primary optimization objective as a relative improvement:
\begin{equation}
    L_{\mathrm{gen}} \leq \alpha_T(k) \cdot L_{\mathrm{base}}
\end{equation}
where $L_{\mathrm{gen}}$ is the latency under the generated policy and $\alpha_T(k)$ is a target fraction.

Furthermore, explicitly defining a specific performance target introduces a
severe secondary challenge: LLM greediness. When instructed solely to maximize a
single foreground metric, LLMs may behave like reward hackers and generate
overly aggressive policies. For example, a scheduler can reduce foreground tail
latency by starving background tasks, delaying system workers, or routing
unrelated tasks into a low-priority path, which may eventually trigger
\texttt{sched\_ext} watchdog panics or system-wide stalls.

To prevent such pathological over-optimization, \system{} formulates scheduler
evaluation as a constrained multi-objective objective rather than a single
foreground-only reward. For latency-sensitive services, let $L_{p99}$ denote the
measured foreground p99 latency under a generated scheduler, and let
$L_{\mathrm{slo}}=\alpha_T(k)L_{\mathrm{base}}$ denote the dynamic latency target
derived from the EEVDF baseline. Let $T_{\mathrm{fg}}$ and $T_{\mathrm{be}}$
denote foreground and background throughput under the generated scheduler, and
let $T_{\mathrm{fg}}^0$ and $T_{\mathrm{be}}^0$ denote the corresponding EEVDF
measurements. Given a required background-retention ratio $\rho_{\mathrm{be}}$,
\system{} defines the normalized foreground violation and background shortfall as
\[
V_{\mathrm{fg}} =
\left[
\frac{L_{p99}-L_{\mathrm{slo}}}{L_{\mathrm{slo}}}
\right]_+,
\qquad
V_{\mathrm{be}} =
\left[
\frac{\rho_{\mathrm{be}}T_{\mathrm{be}}^0 - T_{\mathrm{be}}}
     {T_{\mathrm{be}}^0}
\right]_+,
\]
where $[x]_+=\max(x,0)$. The feedback score used for best-so-far selection and
LLM repair is
\[
\mathcal{J}
=
\lambda_{\mathrm{fg}} V_{\mathrm{fg}}
+
\lambda_{\mathrm{be}} V_{\mathrm{be}}
-
\lambda_{\mathrm{prog}}
\left(
\frac{T_{\mathrm{fg}}}{T_{\mathrm{fg}}^0}
+
\eta \frac{T_{\mathrm{be}}}{T_{\mathrm{be}}^0}
\right).
\]
Lower $\mathcal{J}$ is better. The first two terms penalize violations of the
foreground objective and background-progress requirement, while the last term
breaks ties in favor of candidates that preserve useful system throughput. The
penalty weights are chosen such that
$\lambda_{\mathrm{fg}},\lambda_{\mathrm{be}} \gg \lambda_{\mathrm{prog}}$,
ensuring that throughput rewards cannot compensate for foreground-objective or
background-liveness violations.

For throughput-oriented foreground jobs, \system{} replaces the latency
violation with a normalized foreground-throughput shortfall:
\[
V_{\mathrm{fg}} =
\left[
\frac{T_{\mathrm{slo}}-T_{\mathrm{fg}}}{T_{\mathrm{slo}}}
\right]_+,
\qquad
T_{\mathrm{slo}}=\beta_T(k)T_{\mathrm{fg}}^0 .
\]
A scheduler is accepted only when both $V_{\mathrm{fg}}=0$ and
$V_{\mathrm{be}}=0$ hold. Thus, the score provides a smooth ranking signal for
iterative search, while final acceptance is governed by explicit foreground and
liveness constraints. This design forces the LLM to improve the foreground
workload through interference mitigation rather than trivial starvation.

\subsection{Contract-Driven Progressive Lowering Synthesis}
\label{subsec:synthesis}


A production-ready \texttt{sched\_ext} policy is not a monolithic script; it demands the seamless synchronization of heterogeneous components across the user/kernel boundary. Naively prompting an LLM to generate all source files simultaneously (e.g., as a giant JSON payload) frequently leads to context-window exhaustion, cross-language mismatches, and severe structural hallucinations. 

To overcome this, \system{} abandons end-to-end monolithic generation. Instead, it frames the synthesis as a \textit{Contract-Driven Progressive Lowering} pipeline. This architecture mirrors modern compiler design and microservice governance, decomposing the synthesis into four deterministic, strictly bounded stages.

\textbf{Stage 1: Semantic Extraction and IR Generation (Vertical Lowering).} 
Rather than jumping directly from high-level intents to complex C/Rust syntax, \system{} forces the LLM to first translate the derived diagnostic intent (from Section~\ref{subsec:objective_construction}) into a language-agnostic Scheduler Intermediate Representation (Scheduler IR). As illustrated in Figure~\ref{fig:ir_example}, the IR explicitly formalizes the core logic using structured event-action mappings and pseudo-code (e.g., defining which tasks route to local versus shared dispatch queues, or outlining priority calculation formulas). This progressive lowering prevents the logical fractures that occur when LLMs grapple with algorithmic design and syntactic constraints simultaneously.

\begin{figure}[t]
\centering
\begin{lstlisting}[style=jsonstyle]
{
  "ir_version": "v1",
  "strategy_name": "redis_aggressive_tail_rescue",
  "target": {
    "workload": "redis",
    "match": ["redis-server"],
    "objective": "minimize p99 latency"
  },
  "interferer": {
    "type": "cpu",
    "retention_goal": ">= 85% of EEVDF"
  },
  "event_rule": {
    "event": "enqueue",
    "class": "TARGET",
    "condition": "Redis becomes runnable",
    "action": {
      "queue": "target_dsq",
      "priority_bias": "high",
      "slice": "medium",
      "placement": "low_interference_preferred",
      "starvation_guard": true
    }
  },
  ...
}
\end{lstlisting}
\caption{A compact excerpt of a Scheduler IR generated by \system{} for Redis under CPU interference. The IR captures the target objective, background-retention constraint, one event-action scheduling rule, and safety invariants before Rust+BPF code generation.}
\label{fig:ir_example}
\end{figure}

\textbf{Stage 2: Boundary Explicitization and Contract Locking (Horizontal Decoupling).} 
A minor modification in a BPF dispatch queue rule can trigger cascading mismatches in the Rust-side control logic. To decouple these domains, \system{} extracts the cross-space interaction requirements from the IR and generates a shared \texttt{src/bpf/intf.h} file alongside a \texttt{src/stats.rs} interface. This header defines the exact BPF Maps, \texttt{rodata} configurations, and \texttt{struct\_ops} identifiers. Once generated, these files are frozen and established as an immutable, explicit ``Contract'' between the data plane and the control plane.

\textbf{Stage 3: Contract-Constrained Parallel Realization.} 
With the logical IR and the physical contract established, \system{} fragments the code generation into two parallel LLM threads constrained by a rigid project scaffold:
\begin{itemize}[leftmargin=*]
    \item \textit{The BPF Provider Thread} compiles the IR into the \texttt{main.bpf.c} core callbacks, strictly bound by the data structures defined in the contract.
    \item \textit{The Rust Consumer Thread} translates the IR into the \texttt{main.rs} user-space skeleton, responsible for configuration injection and statistics exporting, possessing only read/write access to the contract's defined structures.
\end{itemize}

Before confronting the strict Linux BPF verifier, \system{} executes a static alignment check. It verifies that the Rust skeleton, the BPF variables, and the \texttt{SCX\_OPS\_DEFINE} macro perfectly align with the frozen contract. Only when both generated planes comply with the \texttt{intf.h} contract is the scheduler passed to the system compiler and subjected to the rigorous memory and execution bounds checks detailed in Table~\ref{tab:verifier-checks}.

\begin{table}[t]
\centering
\small
\caption{Taxonomy of static checks.}
\label{tab:verifier-checks}
\begin{tabular}{p{0.31\linewidth} p{0.65\linewidth}}
\toprule
\textbf{Constraint Area} & \textbf{Specific eBPF / Build Checks} \\
\midrule
\textbf{Skeleton} & Rust/C compilation errors; \texttt{rodata} mapping mismatches; \texttt{struct\_ops} misalignment. \\
\textbf{Memory Safety} & Pointer out-of-bounds; missing NULL-checks; invalid memory dereferences. \\
\textbf{Execution Bounds} & Unbounded loops; instruction limit violations; unreleased spin locks. \\
\textbf{API \& BTF Validity} & Invalid KFunc signatures; BTF type mismatches; unregistered \texttt{scx} helpers. \\
\bottomrule
\end{tabular}
\end{table}

\subsection{Decoupled Dual-Level Optimization and Repair}
\label{subsec:verification_and_feedback}

The true power of \system{} extends beyond mere code generation; it lies in its ability to autonomously navigate the high-dimensional search space of kernel scheduling. Rather than relying on naive, stateless retries that treat all compilation or performance failures equally, \system{} frames the synthesis process as a \textit{Decoupled Dual-Level Optimization and Repair} pipeline. The design of this bifurcated architecture stems from a critical observation of naive LLM agents: standard linear prompting invariably falls into the trap of \textit{defensive accumulation} (compiling compounding, bloated band-aids) or \textit{evasive modification} (destructively bypassing core logic merely to silence the Verifier). 

\system{} recognizes that strategy synthesis and structural repair possess fundamentally distinct mathematical properties. High-level scheduling strategies often exhibit orthogonal benefits; thus, they necessitate an \textit{Evolutionary Memory Pool} to continuously stack and recombine advantageous traits. Conversely, low-level syntactic patches are localized and non-additive. For patches, \system{} abandons recombination and instead utilizes a Directed Acyclic Graph (DAG) strictly as a semantic version control system, enabling precise state check-pointing and rollback to rescue the LLM from evasive dead-ends.

\textbf{1. Outer Loop: Strategy-Level Evolution.}
If a generated scheduler passes all BPF and liveness safety checks but fails to satisfy the performance objective (as defined in Section~\ref{subsec:objective_construction}), continuing to tweak its C/Rust implementation is futile; the underlying strategy is flawed. However, a naive ``strategy refresh''---discarding the current attempt and prompting the LLM from scratch---results in a random, memoryless search. 

\begin{figure}[t]
    \centering
    \includegraphics[width=1\linewidth]{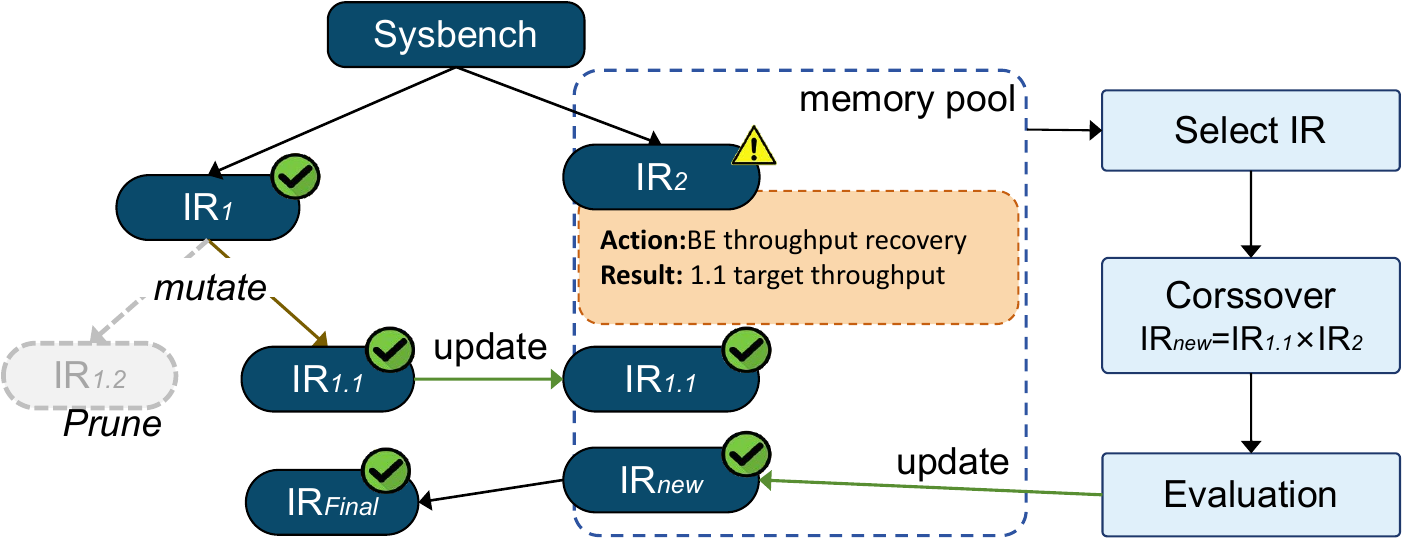}
    \caption{The strategy-level evolutionary process (Outer Loop). The LLM applies mutation and crossover operators to progressively refine the strategies, pruning invalid paths and recombining successful traits to synthesize the final optimal strategy.}
    \label{fig:strategy_evolution}
\end{figure}

Instead, \system{} implements an evolutionary mechanism backed by an \textit{Evolutionary Memory Pool}, as illustrated in Figure~\ref{fig:strategy_evolution}. When a strategy fails the primary target but improves a secondary diagnostic metric (e.g., reducing LLC miss rate without improving overall throughput), its Intermediate Representation (IR) is not discarded. It is archived into the memory pool as a ``sub-optimal but valuable gene.'' 

During subsequent generation attempts, the LLM acts as an evolutionary operator. It retrieves historical fragments from the memory pool to perform either \textit{Mutation} (fine-tuning the existing strategy, such as adapting $\text{IR}_1$ into $\text{IR}_{1.1}$ while pruning invalid branches like $\text{IR}_{1.2}$) or \textit{Crossover} (recombining the core logic of two partially successful historical IRs, such as merging $\text{IR}_{1.1}$ and $\text{IR}_2$). This evolutionary memory ensures that the search process monotonically accumulates scheduling insights rather than blindly guessing.

\textbf{2. Inner Loop: DAG-Structured Repair Engine.}
If the generated code violates static constraints or fails runtime safety checks, the overarching strategy remains valid, but its C/Rust execution is flawed. Traditional LLM-based Automated Program Repair (APR) relies on linear conversational feedback (e.g., $V1 \rightarrow V2 \rightarrow V3$). 

\begin{figure}[t]
    \centering
    \includegraphics[width=1\linewidth]{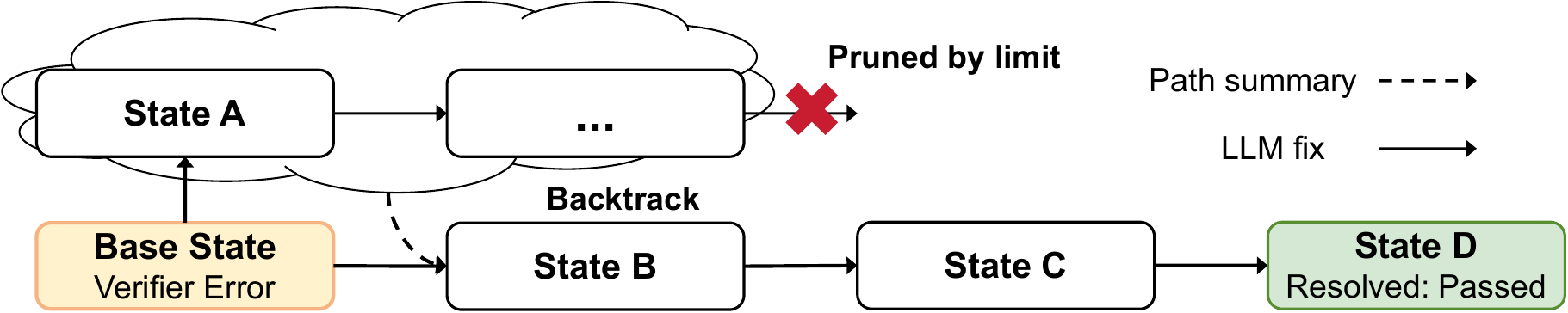}
    \caption{The DAG-structured semantic repair engine navigating eBPF verifier constraints. Nodes represent discrete verification states.}
    \label{fig:dag_repair}
\end{figure}

To conquer this, \system{} abandons the fragile linear prompt history and elevates the repair process into a \textit{DAG-Structured Semantic Repair} framework. As illustrated in Figure~\ref{fig:dag_repair}, instead of a linear chat sequence, \system{} frames program debugging as a systematic state-space search over a Directed Acyclic Graph (DAG) of explicit verification states (nodes) connected by LLM-driven syntactic fixes (edges). Each node in the DAG represents a discrete verification state captured directly from the eBPF verifier. Rather than treating verifier rejections as isolated terminal failures, \system{} tracks the contextual trajectory of these states to detect loops and degenerative code bloat. When an exploration path continues to accumulate compounding verifier errors, the engine actively prunes the active branch to stop futile token consumption. 

A naive rollback at this point would force the LLM to blindly guess alternative fixes, often leading back into the same syntactic traps. To ensure informed and non-overlapping exploration without relying on complex numerical scoring, \system{} encapsulates the entire failed sub-trajectory within a semantic cloud abstraction, automatically extracting a cohesive path summary from the compiled logs of those failed nodes. Crucially, the engine then executes a context rollback, backtracking to a surviving valid ancestor. The extracted path summary is injected directly into the next outward transition edge as a strict negative constraint. By explicitly learning from the summarized failures of the pruned branch, the LLM is forced to break from its previous logic, successfully rationalizing an independent and clean repair path until the trajectory successfully converges onto a fully verified scheduling binary.


\section{Implementation}

We implemented \system{} in Python, Rust, and C. The Python-based synthesis
framework contains 5,485 LOC and implements workflow orchestration, prompt
construction, Scheduler IR generation, LLM invocation, feedback parsing, and
artifact management. The main workflow and feedback modules account for 4,496
LOC. To support long-horizon synthesis, \system{} also includes 989 LOC for the
strategy memory pool and repair memory graph, which record evaluated IRs,
failure states, diagnostic tags, and repair hints for later prompt construction.

To provide a stable generation target, we implemented \texttt{scx\_morphsched}, a
fixed \texttt{sched\_ext} crate inside the upstream \texttt{sched-ext/scx}
repository. The scaffold contains 670 LOC across seven source files, including
307 LOC of Rust, 328 LOC of BPF C, and 35 LOC of shared headers. It handles BPF
skeleton loading, \texttt{struct\_ops} attachment, runtime lifecycle management,
statistics export, and the Rust/BPF shared interface. The LLM is constrained to
generate only four files: \texttt{src/main.rs}, \texttt{src/stats.rs},
\texttt{src/bpf/main.bpf.c}, and \texttt{src/bpf/intf.h}.

\system{} uses an OpenAI-compatible Chat Completions backend configured with
\texttt{gpt-5.5}, temperature 0.2, streaming enabled, a 300-second timeout, and
up to three retries. To ground generated code in the current \texttt{sched\_ext}
API, we built a structured helper database with 58 typed BPF kfunc entries
(476 LOC of JSON). The prompt templates add 305 LOC and combine the workload
objective, Scheduler IR, helper database, and reference scaffold for strategy
generation, IR generation, full-source generation, and repair.

The feedback infrastructure contains 4,722 LOC of scripts and parsers. It runs
generated schedulers, collects runtime and performance results, converts them
into JSON summaries, and feeds deterministic compile/runtime/performance signals
back into the synthesis loop.

\section{Evaluation}
\label{sec:evaluation}

We evaluate \system{} as an end-to-end autonomous synthesis framework. Rather than simply validating a single hand-written policy, our primary objective is to determine whether \system{} can consistently synthesize complete, robust Rust+BPF \texttt{sched\_ext} schedulers that optimize dynamic workload objectives on a real Linux kernel.

Our evaluation methodology spans a comprehensive spectrum of deployment scenarios, ranging from isolated standalone executions(hereafter referred to as Solo) to heavily-interfered colocated environments. In colocated settings, the foreground workload acts as the primary optimization target, while diverse background tasks are introduced to inject controlled micro-architectural contention (e.g., CPU over-subscription, LLC thrashing). 

We assess four diverse foreground targets as Caladan~\cite{fried2020caladan}: Redis, Nginx, and memcached (evaluated by p99 tail latency), alongside the CPU-bound Sysbench (evaluated by aggregate throughput). All performance metrics are strictly normalized against the default Linux EEVDF scheduler under identical conditions, with positive values indicating foreground performance improvement. Crucially, all reported policies are entirely generated by \system{} without any manual source-code intervention.


\subsection{Experimental Setup}
\label{sec:eval-setup}

\paragraph{System and testbeds.}
All experiments run on Ubuntu 24.04 Desktop with Linux
\texttt{6.17.0-29-generic}. \system{} is implemented on top of the
upstream \texttt{sched-ext/scx} repository, and each generated scheduler
is compiled and loaded as a Rust+BPF \texttt{sched\_ext} scheduler
through the eBPF \texttt{struct\_ops} path. The baseline is Linux EEVDF,
i.e., the default scheduler when no \texttt{sched\_ext} scheduler is
attached. Table~\ref{tab:eval-testbeds} summarizes the testbeds.

\begin{table}[t]
\centering
\small
\caption{Evaluation testbeds. VM CPU information is reported by \texttt{lscpu} inside each guest; the underlying virtualization host is not controlled by us.}
\label{tab:eval-testbeds}
\begin{tabular}{lll}
\toprule
\textbf{Platform} & \textbf{CPU resources} & \textbf{Memory} \\
\midrule
Physical
& i9-14900, SMT on, 1 NUMA
& 128~GB \\

VM-4
& 4 vCPUs, Xeon E5-2630 v3
& 16~GB \\

VM-6
& 6 vCPUs, Xeon E5-2630 v3
& 16~GB \\

VM-10
& 10 vCPUs, Xeon E5-2630 v3
& 16~GB \\

VM-16
& 16 vCPUs, Xeon E5-2630 v3, 2 NUMA
& 16~GB \\
\bottomrule
\end{tabular}
\end{table}

\begin{figure*}[t]
    \centering
    \includegraphics[width=1.0\linewidth]{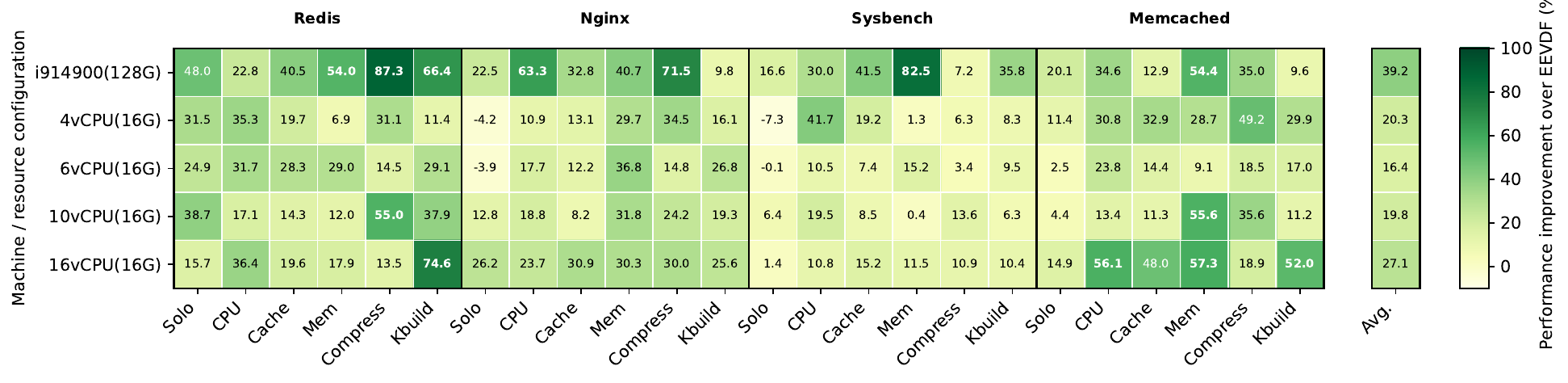}
    \caption{Performance improvement of \system{} over Linux EEVDF
    across machine configurations, foreground workloads, and background
    interferers. For Redis, Nginx, and memcached, improvement is p99
    latency reduction; for Sysbench CPU, improvement is throughput
    increase. Larger values are better.}
    \label{fig:global-eval-heatmap}
\end{figure*}
\paragraph{Workloads.}
Table~\ref{tab:eval-workloads} lists the foreground workloads and
metrics. Each foreground workload is colocated with one of five
background interferers: CPU stress, cache stress, memory stress,
compression, and kernel build. The first three are implemented with
\texttt{stress-ng}; compression repeatedly runs \texttt{pigz}; and kernel
build repeatedly runs Linux build phases~\cite{fried2020caladan}.

\begin{table}[t]
\centering
\small
\caption{Foreground workloads and metrics.}
\label{tab:eval-workloads}
\begin{tabular}{lll}
\toprule
Workload & Benchmark & Metric \\
\midrule
Redis & \texttt{redis-benchmark} & p99 latency \\
Nginx & \texttt{wrk} & p99 latency \\
memcached & \texttt{memtier\_benchmark} & p99 latency \\
Sysbench CPU & \texttt{sysbench} & throughput \\
\bottomrule
\end{tabular}
\end{table}

\paragraph{Protocol and metrics.}
For each machine/workload/interferer configuration, \system{} first runs
the workload under EEVDF to obtain a dynamic baseline, then automatically
generates, compiles, loads, and benchmarks a configuration-specific
\texttt{sched\_ext} scheduler. The process does not involve manual edits
to generated code.

To reduce measurement noise, every reported data point is measured over
three independent runs. Each run starts from a clean scheduler state: any
active \texttt{sched\_ext} scheduler is detached, workload and interferer
processes are restarted, and the generated scheduler is reloaded before
benchmarking. We report the mean across the three runs in all figures and
tables. Error bars are omitted from the heatmap and comparison figures for
readability; however, the repeated runs are used consistently for both EEVDF
and generated schedulers under the same workload, machine, and interferer
setting. For latency-sensitive workloads, we aggregate the p99 latency from
the three benchmark runs; for throughput-oriented workloads, we aggregate
the measured throughput. This paired measurement protocol ensures that each
reported improvement compares \system{} against EEVDF under identical
experimental conditions.

For Redis, Nginx, and memcached, we report p99 latency reduction:
\[
    \frac{L_{\mathrm{EEVDF}} - L_{\mathrm{Gen}}}
         {L_{\mathrm{EEVDF}}} \times 100\%.
\]
For Sysbench CPU, we report throughput improvement:
\[
    \frac{T_{\mathrm{Gen}} - T_{\mathrm{EEVDF}}}
         {T_{\mathrm{EEVDF}}} \times 100\%.
\]
Thus, positive values always indicate better foreground performance than
EEVDF. We also collect diagnostic signals such as CPU utilization,
process statistics, PSI, hardware counters, context switches, CPU
migrations, thread placement, and runnable-to-running latency when
available.

\subsection{End-to-End Performance Improvement}
\label{sec:eval-performance}

Figure~\ref{fig:global-eval-heatmap} shows the end-to-end performance of
\system{} across 120 machine/workload/interferer configurations. Each
cell reports the foreground improvement of the generated scheduler over
Linux EEVDF. For Redis, Nginx, and memcached, improvement denotes p99
latency reduction; for Sysbench CPU, it denotes throughput improvement.

\begin{table*}[t]
\centering
\small
\caption{Average latency improvement over EEVDF for latency-sensitive workloads.
Each cell reports average latency reduction relative to EEVDF; positive values
indicate lower average latency.}
\label{tab:lc-avg-latency-full}
\resizebox{\textwidth}{!}{
\begin{tabular}{lrrrrrrrrrrrrrrrrrr}
\toprule
\multirow{2}{*}{Machine}
& \multicolumn{6}{c}{Memcached}
& \multicolumn{6}{c}{Nginx}
& \multicolumn{6}{c}{Redis} \\
\cmidrule(lr){2-7}
\cmidrule(lr){8-13}
\cmidrule(lr){14-19}
& Solo & CPU & Cache & Mem & Compress & Kbuild
& Solo & CPU & Cache & Mem & Compress & Kbuild
& Solo & CPU & Cache & Mem & Compress & Kbuild \\
\midrule
i9-14900(128G)
& 4.98\% & 0.0\% & 5.9\% & 48.0\% & 19.4\% & 53.7\%
& 3.23\% & 51.5\% & 22.5\% & 94.7\% & 42.2\% & 48.5\%
& 32.93\% & 14.0\% & 15.8\% & 28.1\% & 57.4\% & 22.2\% \\

4vCPU(16G)
&  2.41\% & 7.0\% & 1.1\% & 0.8\% & 38.8\% & 16.4\%
&  -1.44\% & 2.8\% & 10.5\% & 13.4\% & 22.5\% & 11.7\%
&  13.64\% & 28.6\% & 7.3\% & 3.5\% & 16.0\% & 4.3\% \\

6vCPU(16G)
& 0.31\%  & 19.9\% & -1.8\% & {-1.1\%} & 5.2\% & {-8.0\%}
& 0.19\%  & 7.3\% & 14.7\% & 18.6\% & 8.7\% & 14.2\%
& 25.48\% & 20.0\% & 17.8\% & 1.9\% & 6.9\% & 2.5\% \\

10vCPU(16G)
&  {-3.79\%} & {-5.8\%} & 6.5\% & 26.4\% & 4.2\% & 2.0\%
&  21.57\% & 19.6\% & 6.6\% & 7.5\% & 29.5\% & 25.8\%
& 8.85\% & 3.7\% & 8.7\% & {-8.7\%} & {-4.65\%} & {-0.8\%} \\

16vCPU(16G)
&  3.80\% & 28.4\% & 18.4\% & 18.8\% & 9.6\% & 23.4\%
&  11.73\% & 22.8\% & 2.4\% & 2.1\% & 18.5\% & 4.8\%
&  3.17\% & 16.4\% & 1.0\% & 11.7\% & 9.6\% & 11.3\% \\
\bottomrule
\end{tabular}
}
\end{table*}

\begin{table}[t]
\centering
\small
\caption{Background CPU throughput retention under generated schedulers. Each cell reports stress-ng CPU throughput relative to EEVDF; larger is better.}
\label{tab:be-cpu-throughput-retention}
\begin{tabular}{lcccc}
\toprule
& Redis & Nginx & memcached & Sysbench \\
\midrule
BE retention
& 98.1\% & 89.9\% & 116.3\% & 100.9\% \\
\bottomrule
\end{tabular}
\vspace{2pt}
\begin{minipage}{0.95\linewidth}
\footnotesize
\end{minipage}
\end{table}

\begin{table}[t]
\centering
\small
\caption{Iteration outcomes across selected workload/interferer cases. For each case, we select an accepted run when available; otherwise, we use the lowest-score fallback run.}
\label{tab:selected-case-iteration-outcomes}
\begin{tabular}{lr}
\toprule
Metric & Value \\
\midrule
Compile \& Verify errors & 12 \\
Performance not met & 33 \\
Accept iterations & 24 \\
\midrule
Total iterations & 69 \\
Avg. iterations per case & 2.87 \\
\bottomrule
\end{tabular}
\end{table}

Overall, \system{} improves foreground performance in all evaluated
configurations. Across the 120 cases, the average improvement is 26.5\%,
with a maximum improvement of 87.3\%. This result shows that \system{}
can synthesize complete \texttt{sched\_ext} schedulers that not only
compile and run safely, but also improve workload-specific objectives on
the real Linux scheduling path.

The gains vary across workloads, which confirms that scheduler synthesis
is workload-dependent. Latency-sensitive services benefit substantially:
Redis, memcached, and Nginx improve by 32.2\%, 30.4\%, and 26.9\% on
average, respectively. These workloads are sensitive to wakeup and
dispatch delay, allowing generated policies such as target wakeup boost,
priority bias, and soft core protection to reduce tail latency. Sysbench
CPU also improves in every configuration, but with a lower average gain
of 16.3\%, reflecting its different optimization target: stable CPU
placement and reduced interference rather than request-level tail
latency.

The improvement is also consistent across machine configurations. The
physical i9-14900 machine achieves an average improvement of 40.6\%,
while the 4-vCPU, 6-vCPU, 10-vCPU, and 16-vCPU virtual machines achieve
22.8\%, 18.5\%, 20.7\%, and 29.7\% average improvement, respectively.
Thus, \system{} does not only optimize for a single testbed; it can
generate effective schedulers across both physical and virtualized
environments.

To complement the p99 results in Figure~\ref{fig:global-eval-heatmap},
Table~\ref{tab:lc-avg-latency-full} reports average latency reduction for
the latency-sensitive services. Average latency is a stricter secondary view:
although \system{} improves p99 latency in all heatmap configurations, a few
average-latency cases regress, as marked in red.

Finally, the benefit depends on the interference type. Memory and
compression interference show the largest average gains, while cache
interference is comparatively harder to mitigate. This is expected:
memory- and compression-heavy interferers create scheduling opportunities
through CPU placement and foreground protection, whereas cache pollution
is only indirectly controllable by the scheduler. These results answer
RQ1: \system{} can automatically synthesize workload-specific
\texttt{sched\_ext} schedulers that consistently outperform EEVDF under
colocated workloads.

\subsection{Comparison with Upstream \texttt{scx} Schedulers}
\label{sec:eval-scx-comparison}

We next compare \system{} with representative hand-written schedulers
from the upstream \texttt{scx} repository. This experiment separates the
benefit of using \texttt{sched\_ext} from the benefit of synthesizing a
workload-specific policy. We select five upstream schedulers that run
unmodified on our testbed: \texttt{scx\_lavd}, \texttt{scx\_rusty},
\texttt{scx\_tickless}, \texttt{scx\_bpfland}, and 
\texttt{scx\_flash}. These schedulers represent existing manually
engineered \texttt{sched\_ext} policies, while \system{} generates a
scheduler for the given workload/interferer objective.

\begin{figure*}[t]
    \centering
    \includegraphics[width=1.0\linewidth]{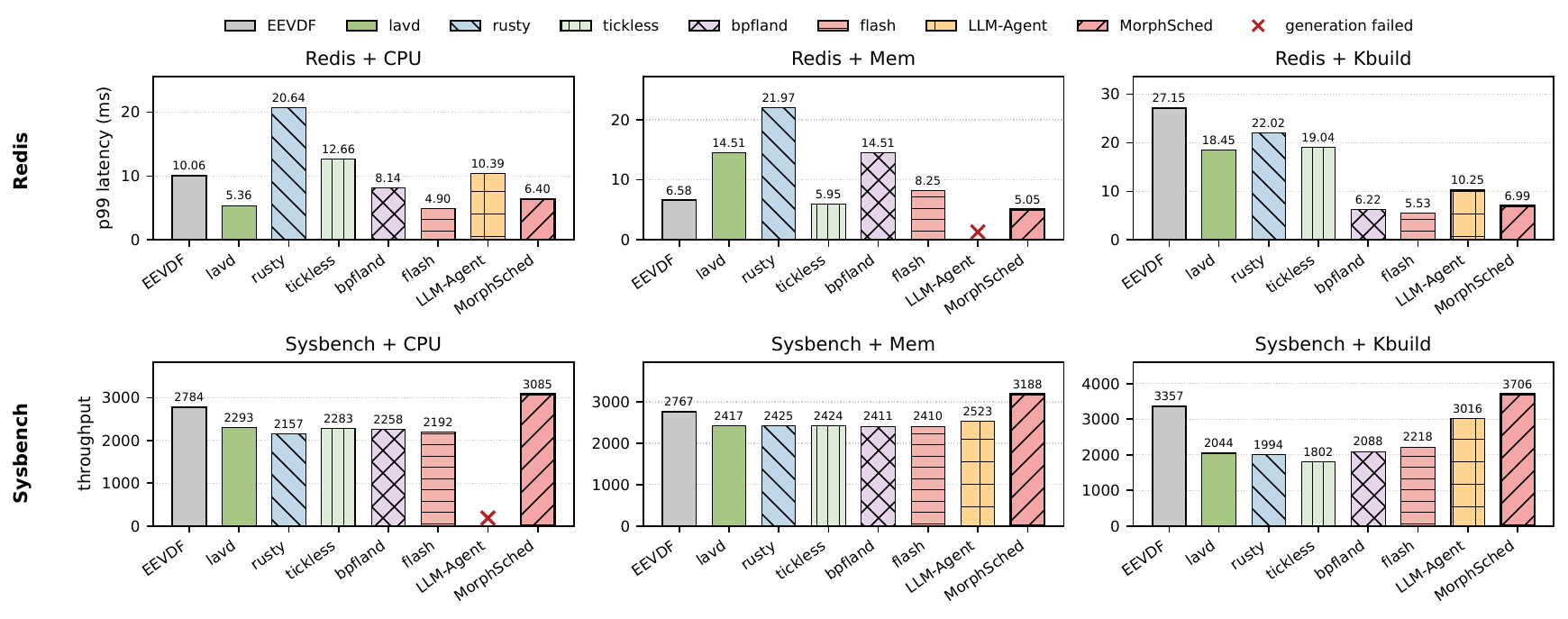}
    \caption{Comparison with upstream \texttt{scx} schedulers on selected
    workload/interferer cases. The first row reports Redis p99 latency
    in milliseconds, where lower is better. The second row reports
    Sysbench CPU throughput, where higher is better. EEVDF is the Linux
    default scheduler.}
    \label{fig:scx-comparison}
\end{figure*}

Figure~\ref{fig:scx-comparison} shows the comparison on six
representative colocated cases. The first row uses Redis, representing a
latency-sensitive service, and reports p99 latency in milliseconds
(lower is better). The second row uses Sysbench CPU, representing a
throughput-oriented foreground job, and reports throughput (higher is
better). We include EEVDF directly as the Linux default baseline to make
the absolute performance differences visible.

For Redis, \system{} consistently improves over EEVDF and is competitive
with the best upstream \texttt{scx} schedulers. Under CPU interference,
\system{} reduces p99 latency from 10.06~ms to 6.40~ms, although
\texttt{scx\_flash} and \texttt{scx\_lavd} perform better in this case.
Under memory interference, \system{} achieves the lowest p99 latency
among all compared schedulers, reducing p99 latency from 6.58~ms to
5.05~ms. Under kernel-build interference, \system{} reduces Redis p99
latency from 27.15~ms to 6.99~ms, close to the best hand-written
scheduler in this case. These results show that \system{} can synthesize
service-oriented policies that are competitive with manually designed
\texttt{sched\_ext} schedulers.

The Sysbench results highlight a different effect. In all three
Sysbench cases, \system{} achieves the highest throughput, while the
upstream \texttt{scx} schedulers reduce throughput relative to EEVDF.
For example, under kernel-build interference, \system{} improves
throughput from 3357 to 3706, whereas the upstream \texttt{scx}
schedulers range from 1802 to 2218. This suggests that a general-purpose
hand-written policy may not match the objective of a CPU-bound foreground
job, while \system{} can generate a policy that better preserves CPU
placement and foreground progress.

Overall, the comparison shows that \system{} is not merely benefiting
from replacing EEVDF with any \texttt{sched\_ext} scheduler. Existing
\texttt{scx} schedulers can perform well in some cases, but their
behavior varies across workload objectives. By synthesizing a scheduler
for each workload/interferer configuration, \system{} achieves
competitive Redis tail latency and consistently higher Sysbench
throughput in the selected cases.

\subsection{Generated Scheduler Code Complexity}
\label{sec:eval-loc}

We next examine the size of generated schedulers. The goal is not to use
LOC as a performance metric, but to quantify the engineering footprint of
the code synthesized by \system{}. Unlike approaches that generate only a
small policy function, \system{} produces complete multi-file Rust+BPF
\texttt{sched\_ext} schedulers, including user-space control logic, BPF
scheduling callbacks, statistics, and shared interfaces.

\begin{figure}[t]
    \centering
    \includegraphics[width=1.0\linewidth]{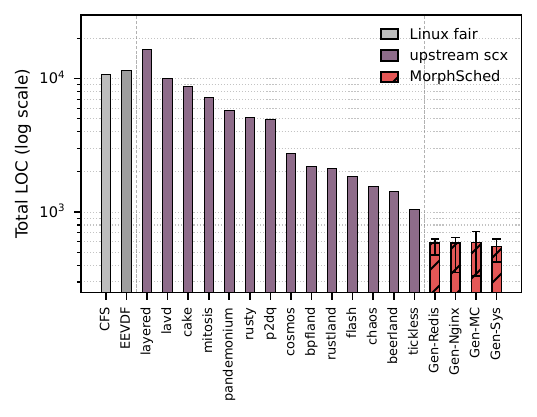}
    \caption{Code size comparison between Linux fair schedulers,
    upstream \texttt{scx} schedulers, and \system{}-generated schedulers.
    CFS and EEVDF count \texttt{kernel/sched/fair.c}; upstream
    \texttt{scx} bars count scheduler source files; \system{} bars report
    median generated LOC for each target workload with min--max ranges.
    The y-axis uses a log scale.}
    \label{fig:loc-comparison}
\end{figure}

Figure~\ref{fig:loc-comparison} compares \system{} with Linux fair
schedulers and upstream \texttt{scx} schedulers. For CFS and EEVDF, we
count the primary fair-scheduler implementation file,
\texttt{kernel/sched/fair.c}. For upstream \texttt{scx} schedulers, we
count each scheduler's source files. For \system{}, we report the median
LOC of final generated schedulers for each target workload, with
min--max ranges across machine/interferer configurations.

The generated schedulers are substantially more compact than most
hand-written schedulers while still being complete implementations. CFS
and EEVDF each exceed 10K LOC in \texttt{fair.c}, and upstream
\texttt{scx} schedulers range from roughly 1K LOC to more than 16K LOC.
In contrast, \system{}-generated schedulers are in the hundreds of LOC
for each workload. The generated size varies across Redis, Nginx,
memcached, and Sysbench, reflecting workload-specific policy structure,
but remains within a compact range because the scaffold fixes the
project layout and cross-file interfaces.

These results answer RQ3: \system{} synthesizes nontrivial, complete
Rust+BPF scheduler implementations rather than small snippets, while
keeping the generated code compact enough for automated compilation,
repair, and feedback-driven iteration.

\subsection{Scheduling Behavior Analysis}
\label{sec:eval-behavior}

We finally analyze the runtime behavior of generated schedulers. The
purpose of this experiment is not to show that every low-level metric
improves monotonically, but to understand what tradeoffs \system{} makes
when optimizing the foreground objective. The metrics in this section are
therefore diagnostic signals rather than independent optimization
targets.

\begin{figure}[t]
    \centering
    \includegraphics[width=1.0\linewidth]{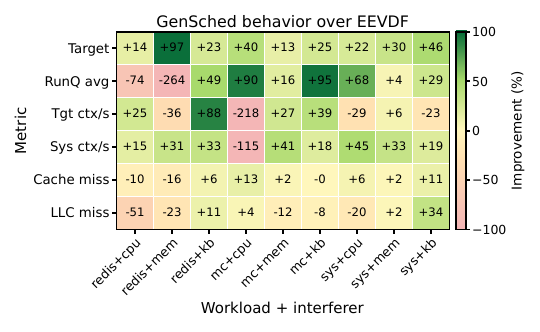}
    \caption{Scheduling behavior of \system{} relative to EEVDF on
    representative colocated workloads. Each cell reports the improvement
    of \system{} over EEVDF. For the target metric, positive values denote
    p99 latency reduction for Redis and memcached, and throughput
    improvement for Sysbench. For the other rows, positive values denote
    lower metric values under \system{}.}
    \label{fig:rq4-behavior}
\end{figure}

Figure~\ref{fig:rq4-behavior} compares \system{} with EEVDF on Redis,
memcached, and Sysbench under CPU, memory, and kernel-build
interference. We orient all metrics so that positive values indicate
better behavior. The first row reports the foreground objective: p99
latency reduction for Redis and memcached, and throughput improvement for
Sysbench. The remaining rows report diagnostic scheduling and hardware
signals, including runnable-to-running delay, target context-switch rate,
system context-switch rate, cache miss rate, and LLC miss rate.

The target metric improves in all selected cases. Redis improves by
47\%, 97\%, and 23\% under CPU, memory, and kbuild interference,
respectively. memcached improves by 40\%, 13\%, and 54\%, while Sysbench
throughput improves by 22\%, 30\%, and 46\%. This confirms that the
generated schedulers optimize the foreground workload across both
latency-sensitive services and CPU-bound foreground jobs.

The diagnostic metrics show that these improvements come with
workload-dependent tradeoffs. Runnable average latency improves in most
memcached and Sysbench cases, but worsens for Redis under CPU and memory
interference. Similarly, target context-switch rate does not decrease
monotonically. This is expected: \system{} optimizes the workload-level
objective derived from the EEVDF baseline, rather than independently
minimizing every low-level counter. A generated policy may trade
additional wakeup or context-switch activity for better request-level
tail latency or higher foreground throughput.

More consistent improvements appear in system-level and cache-related
metrics. \system{} reduces system context-switch rate in most cases, and
cache miss rate improves in eight of the nine selected cases. LLC miss
rate also improves in six cases. These trends suggest that generated
policies often reduce resource interference, likely through placement,
dispatch bias, or bounded control of background work. At the same time,
the mixed behavior of runnable delay and target context switches shows
that there is no single low-level metric that explains all performance
gains.

Overall, these results answer RQ4: \system{} synthesizes schedulers that
make workload-specific tradeoffs. The generated schedulers consistently
improve the foreground objective and often reduce system context-switch
activity and cache interference, but some target-level scheduling metrics
may worsen in individual cases. This reinforces the need for a
feedback-driven synthesis loop that evaluates complete workload outcomes
rather than optimizing a fixed low-level heuristic.

\subsection{Generalization Across LLM Backends}
\label{subsec:model_comparison}

\begin{figure}[t]
    \centering
    \includegraphics[width=1.0\linewidth]{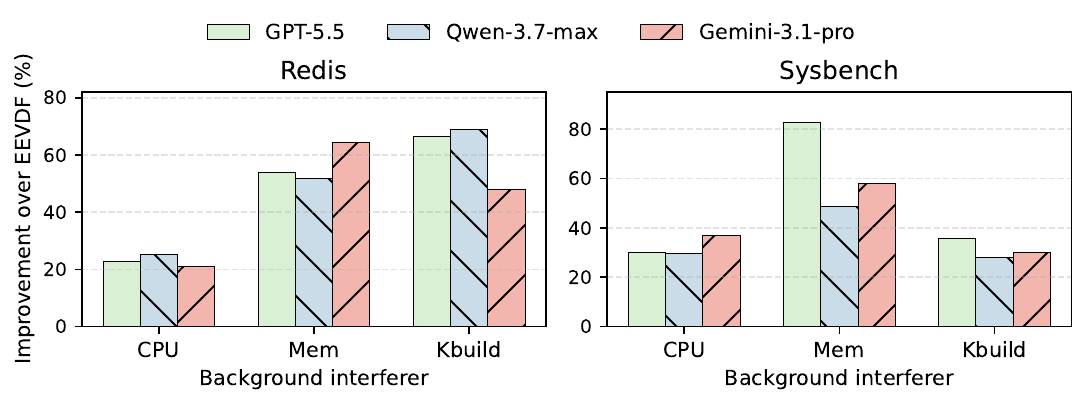}
    \caption{Performance of \system{} when instantiated with different LLM
    backends. Redis reports p99 latency reduction over EEVDF, while Sysbench
    reports throughput improvement over EEVDF. Larger is better.}
    \label{fig:model_comparison}
\end{figure}

To understand whether \system{} depends on a specific LLM backend, we repeat the
synthesis workflow with three different models: GPT-5.5, Qwen-3.7-max, and
Gemini-3.1-pro. We evaluate two representative foreground workloads, Redis and Sysbench,
under three background interferers: CPU, memory, and kernel build.

Figure~\ref{fig:model_comparison} shows that \system{} consistently synthesizes
beneficial schedulers across all three models. For Redis, all generated
schedulers reduce p99 latency over EEVDF across the three interference settings.
The best model varies by case: Qwen3.7-max achieves the largest gain under
Kbuild interference (69.05\%), Gemini performs best under memory interference
(64.46\%), and the three models are close under CPU interference
(21.05--25.08\%). This suggests that the structured workflow, rather than a
single model-specific behavior, is responsible for producing useful scheduling
policies.

The same trend holds for Sysbench. All three models improve throughput over
EEVDF in every case, with GPT~5.5 reaching the largest gain under memory
interference (82.50\%), Gemini performing best under CPU interference
(37.05\%), and GPT~5.5 leading under Kbuild interference (35.80\%). Although the
absolute gains differ across models, no backend dominates all cases. These
results indicate that \system{} is not tightly coupled to one proprietary model:
the IR-guided generation, constrained source format, and feedback-driven repair
loop provide enough structure for multiple LLMs to synthesize effective
\texttt{sched\_ext} schedulers.

\subsection{Ablation}

\begin{figure}[t]
    \centering
    \includegraphics[width=1.0\linewidth]{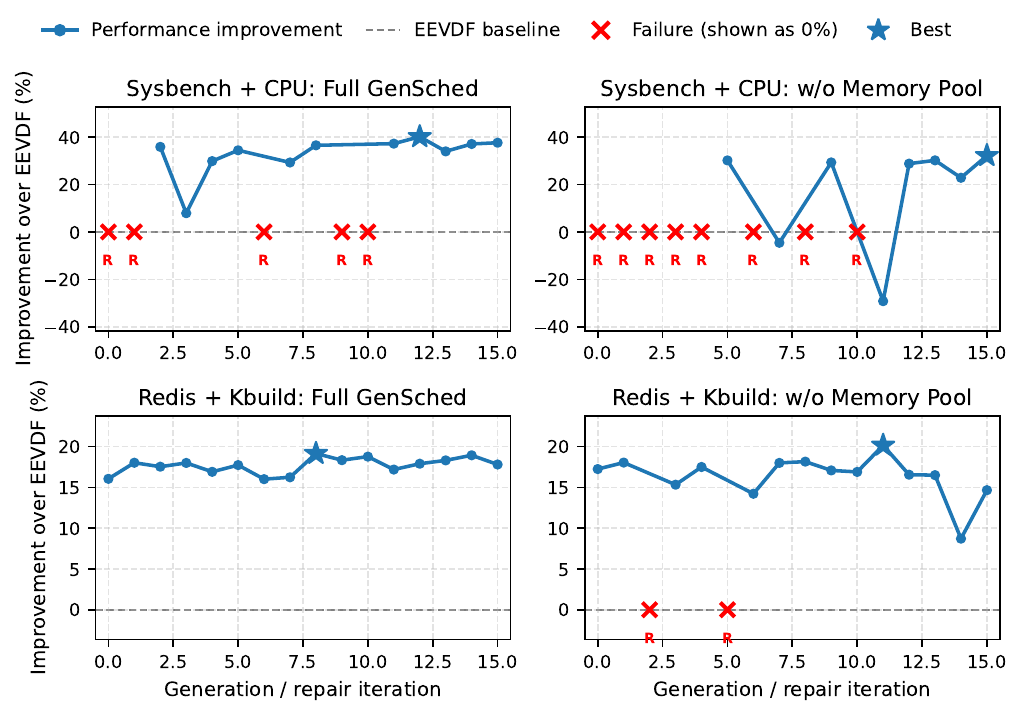}
    \caption{Ablation of \system{}'s history-aware dual-level search.
    Each point reports foreground improvement over EEVDF at one
    generation/repair iteration. Failed iterations are plotted at 0\%
    improvement, corresponding to falling back to EEVDF-equivalent behavior.
    Stars mark the best candidate in each run.}
    \label{fig:ablation-memory}
\end{figure}

Figure~\ref{fig:ablation-memory} isolates the contribution of the
history-aware search components in \system{}. The ablated variant preserves the
same workload objective construction, Scheduler IR generation, compile/runtime/
performance feedback, strategy refresh, and implementation repair. However, it
disables both parts of the dual-level search memory: the strategy-level
evolutionary memory pool and the repair-level DAG memory. Thus, the ablated
system still receives feedback, but each refresh or repair is driven primarily
by the current failure rather than by accumulated search history.

The results show that the history-aware mechanism improves long-horizon
synthesis stability. In the Sysbench+CPU case, full \system{} quickly enters a
high-performance region and remains mostly around 30--40\% improvement after the
initial invalid candidates. Without the dual-level memory, the search produces
more failed iterations and larger performance swings, including a severe
regression before recovering. This suggests that retaining prior repair states
and strategy outcomes helps the LLM avoid repeating unsafe implementation
directions and preserve useful scheduling structure across iterations.

The Redis+Kbuild case exhibits a smaller performance range, but the same trend.
Both variants eventually produce useful schedulers, yet full \system{} reaches a
stable band earlier and shows fewer invalid iterations. The ablated system still
benefits from IR-guided generation and real-system feedback, but lacks the
search continuity provided by the strategy memory pool and repair graph.
Overall, the ablation indicates that the dual-level history mechanism is not
merely additional prompt context; it provides reusable search state that improves
the robustness and efficiency of iterative scheduler synthesis.

\subsection{Case Study: Redis under CPU Interference}

We use Redis under CPU interference to illustrate the concrete scheduler
synthesized by \system{}. This case exposes a common reward-hacking risk:
a greedy policy could reduce Redis tail latency by starving
\texttt{stress-ng} workers. \system{} instead requires foreground improvement
only under a background-retention constraint.

Under EEVDF, Redis achieves 1.58M requests/s with 1.460 ms average latency and
14.087 ms p99 latency, while the CPU interferer achieves 40,743 bogo ops/s.
\system{} synthesizes a ``Safe Redis latency scheduler'' that classifies only
\texttt{redis-server} as the target, only \texttt{stress-ng}/\texttt{stress\_cpu}
workers as background interferers, and all remaining tasks as \texttt{OTHER}.
With this scheduler, Redis p99 latency drops to 10.879 ms, a 22.8\% reduction,
and average latency drops to 1.256 ms, a 14.0\% reduction. Redis throughput
slightly increases to 1.59M requests/s, while the CPU interferer retains 89.2\%
of its EEVDF throughput.

Algorithm~\ref{alg:redis-case-policy} summarizes the generated enqueue policy.
The generated BPF implementation lowers the Scheduler IR into three task classes:
target Redis tasks receive a bounded wakeup bonus, CPU interferers receive a
bounded penalty only while Redis is recently active, and \texttt{OTHER} tasks use
a shared non-starving dispatch path.

\begin{algorithm}[t]
\small
\caption{Generated Redis-aware enqueue policy}
\label{alg:redis-case-policy}
\begin{algorithmic}[1]
\Procedure{Enqueue}{$task$}
    \If{$task$ matches \texttt{redis-server}}
        \State classify $task$ as \textsc{Target}
        \State apply bounded wakeup bonus
        \State mark Redis as recently active
    \ElsIf{$task$ matches \texttt{stress-ng} \textbf{and} Redis is recently active}
        \State classify $task$ as \textsc{Background}
        \State apply bounded background penalty
    \Else
        \State classify $task$ as \textsc{Other}
        \State use shared non-starving dispatch path
    \EndIf
\EndProcedure
\end{algorithmic}
\end{algorithm}

This case shows that \system{} does not simply elevate the foreground workload
unconditionally. Instead, it synthesizes a bounded interference-mitigation policy:
Redis receives short-lived wakeup protection, background CPU work continues to
make progress, and system/client tasks keep an explicit fallback path.

\section{Discussion}
\label{sec:discussion}

\textbf{Workload-specific policies.}
\system{} does not aim to replace EEVDF with a single universal scheduler.
Instead, it synthesizes policies for a specific workload, interferer, and
machine setting. As shown in the behavior analysis, generated schedulers may
improve the foreground objective while making different tradeoffs in runnable
latency, context switches, or cache behavior. This reinforces our design choice:
low-level counters are useful diagnostics, but the workload-level objective
remains the optimization target.

\textbf{Validation boundary.}
\system{} validates generated schedulers through the Linux build system, the
eBPF verifier, \texttt{sched\_ext} attach checks, runtime monitoring, and
workload execution. These checks are sufficient for automated exploration on the
real Linux scheduling path, but they are not a formal correctness proof. Stronger
static analysis, longer stress testing, and deployment-specific guardrails would
be needed before using generated policies in production.

\textbf{Model and API dependence.}
Generation quality depends on the backend LLM, prompt context, helper database,
and kernel version. This is especially relevant for \texttt{sched\_ext}, whose
helper signatures, callbacks, and reference schedulers continue to evolve.
\system{} mitigates this with a fixed scaffold, structured helper context, and
compile/runtime feedback, but future systems should automatically extract API
context from the target kernel and \texttt{scx} repository.

\textbf{Generality.}
Our evaluation focuses on one foreground target colocated with one background
interferer. This setting exposes important scheduling tradeoffs, but broader
deployments may involve multiple foreground services, cgroups, NUMA effects,
I/O-heavy workloads, or interacting SLOs. Extending \system{} to multi-objective
and multi-tenant scheduling is a natural next step.


\section{Related Work}

\subsection{Latency- and Interference-Aware Scheduling}

Prior work has directly managed low-level scheduling and resource interference
to protect latency-sensitive workloads. Heracles~\cite{lo2015heracles} and
PARTIES~\cite{chen2019parties} use dynamic resource isolation and feedback
control to protect latency-critical services while improving utilization.
Shinjuku~\cite{kaffes2019shinjuku} targets microsecond-scale tail latency with
centralized dispatching and frequent preemption, while
Caladan~\cite{fried2020caladan} shows that interference can change at
microsecond timescales and uses fast monitoring and core allocation beyond the
default Linux scheduling path. Other systems design specialized policies for
particular deployment settings or interference sources, such as asymmetric
scheduling in edge environments or SMT-aware CPU allocation
~\cite{jia2022achieving, pi2022holmes, kim2026smtcheck}. In contrast to these
handcrafted policies, our work treats scheduler logic as the object of
generation and search, producing deployable scheduler variants that are
evaluated under realistic co-running workloads.

\subsection{Adaptive and Learning-Based Scheduling}

Adaptive and learning-based scheduling seeks to move beyond fixed handcrafted
policies under dynamic workloads. Prior work has applied machine learning or
reinforcement learning to cluster scheduling, kernel resource management, and
OS subsystem decisions~\cite{mao2019learning, fingler2023towards,
doudali2019kleio}, but such methods often require predefined features, state
and action spaces, and careful control of training cost, inference overhead,
stability, and kernel integration. Extensible scheduling frameworks such as
ghOSt and sched\_ext/scx lower the barrier to implementing custom policies
~\cite{humphries2021ghost, schedext}, and ASA uses a
learned router to select among expert schedulers~\cite{wang2025mixture}. In
contrast, our work generates scheduler logic itself, producing deployable
scheduler variants that are evaluated under realistic workloads.

\subsection{Generative Methods for Systems Optimization}

Generative methods have recently been applied to a range of systems software
development and optimization tasks, including program synthesis, code
generation, formal specification generation, verification support,
configuration tuning, and performance diagnosis. Traditional program synthesis
approaches, such as SyGuS, Sketch, and Rosette, typically rely on DSLs, formal
constraints, or restricted search spaces to generate programs satisfying given
specifications~\cite{padhi2023sygus, solar2008program, torlak2014lightweight}.
Recent LLM-based systems work further applies generative models to file-system
specification generation, OS verification tasks, Rust proof assistance,
performance suggestions, and large-scale system diagnosis~\cite{liu2026sharpen,
ma2025specgen, li2026osvbench, yang2025autoverus, stoica2024if, park2025principles}. These efforts reduce manual
effort in system development and verification, but mostly generate code
fragments, specifications, configurations, or recommendations rather than
deployable system policies. In contrast, our work applies generation to OS
scheduler construction, producing scheduler variants that can be compiled,
deployed, and evaluated under realistic co-running workloads.

\section{Conclusion}
\label{sec:conclusion}

This paper presents \system{}, a feedback-driven system for generative
Linux scheduler synthesis with LLMs and \texttt{sched\_ext}. \system{}
uses workload descriptions, \texttt{sched\_ext} API context, reference
scaffolds, and compile/runtime/performance feedback to synthesize
complete Rust+BPF schedulers rather than small policy snippets. By
constructing dynamic objectives from EEVDF baselines and iterating over
implementation repair and strategy refresh, \system{} turns scheduler
design into an empirical generate--validate--optimize loop.

Our evaluation shows that \system{} can generate workload-specific
schedulers that improve foreground performance across colocated
workloads. Across 120 machine/workload/interferer configurations,
\system{} improves over EEVDF in all cases, with an average improvement
of 26.5\% and a maximum improvement of 87.3\%. The generated schedulers
are competitive with representative upstream \texttt{scx} schedulers,
remain compact while containing complete Rust/BPF components, and exhibit
workload-dependent scheduling behavior. These results suggest that LLMs,
when constrained by real system feedback and the \texttt{sched\_ext}
interface, can become practical tools for exploring and synthesizing
Linux scheduling policies.

\bibliography{software}
\bibliographystyle{plain}
\end{document}